\newif\ifARXIV
\ARXIVfalse
\ARXIVtrue

\ifARXIV
\documentclass[sigconf,nonacm]{acmart}
\else
\documentclass[sigconf,review]{acmart}
\fi
\newif\ifDEBUG
\DEBUGtrue
\DEBUGfalse

\usepackage{xcolor}
\usepackage{tcolorbox}
\usepackage{soul} 
\usepackage{xspace}
\usepackage{tabularx}
\usepackage{booktabs}
\usepackage{makecell}
\usepackage{array}
\usepackage{graphicx}
\usepackage{caption}
\usepackage[skip=2pt]{caption}
\ifDEBUG
    \newcommand{\JD}[1]{\textcolor{red}{[\textbf{JD:} #1]}}
    \newcommand{\GKT}[1]{\textcolor{purple}{[\textbf{GKT:} #1]}}
    \newcommand{\TS}[1]{\textcolor{blue}{[\textbf{TS:} #1]}}
    \newcommand{\WJ}[1]{\textcolor{cyan}{[\textbf{WJ:} #1]}}
    \newcommand{\KL}[1]{\textcolor{teal}{[\textbf{KL:} #1]}}
    \newcommand{\EP}[1]{\textcolor{orange}{[\textbf{EP:} #1]}}
    \newcommand{\HP}[1]{\textcolor{orange}{[\textbf{HP:} #1]}}
    \newcommand{\TODO}[1]{\hl{#1}}
    \newcommand{\todo}[1]{%
      \begin{tcolorbox}[colback=yellow!10!white,
                        colframe=red!80!black,
                        boxrule=0.8pt,
                        arc=2pt,
                        left=4pt,
                        right=4pt,
                        top=2pt,
                        bottom=2pt]
        \textbf{TODO:} #1
      \end{tcolorbox}%
    }
\else
    \newcommand{\JD}[1]{}
    \newcommand{\GKT}[1]{}
    \newcommand{\TS}[1]{}
    \newcommand{\WJ}[1]{}
    \newcommand{\KL}[1]{}
    \newcommand{\EP}[1]{}
    \newcommand{\HP}[1]{}
    \newcommand{\TODO}[1]{}
    \newcommand{\todo}[1]{}
\fi

\newcommand{\eg}{\textit{e.g.},\xspace}
\newcommand{\etal}{\textit{et al.}\xspace}

\usepackage{hyperref}

\usepackage{cleveref}
\crefformat{section}{\S#2#1#3}
\crefname{figure}{Figure}{Figures}
\crefname{table}{Table}{Tables}
\crefname{theorem}{Theorem}{Theorems}
\crefname{thm}{Theorem}{Theorems}
\crefname{lemma}{Lemma}{Lemmata}
\crefname{equation}{Eqt.}{Eqts.}
\crefformat{Grammar}{Grammar #1}
\crefname{appendix}{Appendix}{Appendices}
\crefname{listing}{Listing}{Listings}

\usepackage{listings}
\lstdefinelanguage{json}{
  basicstyle=\ttfamily\small,
  morestring=[b]",
  showstringspaces=false,
  breaklines=true
}
\usepackage{hyperref}
\usepackage{url}

\AtBeginDocument{%
    }

\setcopyright{none}
\copyrightyear{2025}
\acmYear{2025}
\acmDOI{XXXXXXX.XXXXXXX}
\acmConference[ICSE '26]{International Conference on Software Engineering}{April 12--18, 2026}{Rio De Janeiro, Brazil}
\acmISBN{XXX-X-XXXX-XXXX-X/2025/11}

\usepackage{cleveref}
\crefrangeformat{section}{\S#3#1#4--#5#2#6}

\begin{document}

\title{AgentHub: A Registry for Discoverable, Verifiable, and Reproducible AI Agents}

\author{Erik Pautsch}
\authornote{Erik Pautsch and Tanmay Singla contributed equally to this work.}
\authornote{Emails: \href{mailto:epautsch@luc.edu}{epautsch@luc.edu}, \href{mailto:klaufer@luc.edu}{klaufer@luc.edu}, \href{mailto:gthiruvathukal@luc.edu}{gthiruvathukal@luc.edu}}
\author{Konstantin Läufer}
\author{George K. Thiruvathukal}

\affiliation{%
  \institution{Loyola Univ.\ Chicago, USA}
  \country{}
}

\author{Tanmay Singla}
\authornotemark[1]
\authornote{Emails: \href{mailto:singlat@purdue.edu}{singlat@purdue.edu}, \href{mailto:kumar565@purdue.edu}{kumar565@purdue.edu}, \href{mailto:peng397@purdue.edu}{peng397@purdue.edu}, \href{mailto:davisjam@purdue.edu}{davisjam@purdue.edu}}

\author{Parv Kumar}
\author{Huiyun Peng}
\author{James C. Davis}

\affiliation{%
  \institution{Purdue University, USA}
  \country{}
}

\author{Wenxin Jiang}
\authornote{\href{mailto:wenxin@socket.dev}{wenxin@socket.dev}}
\affiliation{%
  \institution{Socket Inc.}
  \country{USA}
}

\author{Behnaz Hassanshahi}
\authornote{\href{mailto:behnaz.hassanshahi@oracle.com}{behnaz.hassanshahi@oracle.com}}
\affiliation{%
  \institution{Oracle Labs}
  \country{Australia}
}

\renewcommand{\shortauthors}{Pautsch, Singla, \etal}

\begin{abstract}
LLM-based agents are rapidly proliferating, yet the infrastructure for discovering, evaluating, and governing them remains fragmented compared to mature ecosystems like software package registries (\eg npm) and model hubs (\eg Hugging Face).
Existing efforts typically address naming, distribution, or protocol descriptors, but stop short of providing a registry layer that makes agents discoverable, comparable, and governable under automated reuse.

We present \textit{AgentHub}, a registry layer and accompanying research agenda for agent sharing that targets discovery and workflow integration, trust and security, openness and governance, ecosystem interoperability, lifecycle transparency, and capability clarity with evidence.
We describe a reference prototype \JD{prototype? architecture? `prototype' means we've built out the core features fully} that implements a canonical manifest with publish-time validation, version-bound evidence records linked to auditable artifacts, and an append-only lifecycle event log whose states are respected by default in search and resolution.
We also provide initial discovery results using an LLM-as-judge recommendation pipeline, showing how structured contracts and evidence improve intent-accurate retrieval beyond keyword-driven discovery.
AgentHub aims to provide a common substrate for building reliable, reusable agent ecosystems.

\end{abstract}

\maketitle

\section{Introduction}
\label{sec:Introduction}

LLM-based agents are rapidly entering workflows, from scientific discovery~\cite{gottweis_towards_2025} to software engineering~\cite{zhang2024autocoderover}. 
Unlike static software packages or pretrained models~\cite{JiangPTMReuse}, \emph{agents} act with autonomy, compose tools dynamically, evolve (self-refine) over time, and can operate at scale~\cite{GoogleCloudAIAgents,He2025LLMMultiAgent}. 
We believe these attributes necessitate a new approach to sharing and composing the associated artifacts.
As agent adoption grows, the lack of suitable infrastructure risks limiting both research progress and real-world impact.

In designing a registry for agent sharing, we can learn from the ecosystems for earlier kinds of software.
Conventional registries such as PyPI, npm, and Maven Central show the value of structured metadata, dependency graphs, and signed provenance~\cite{Zimmermann2019,schorlemmer2024signing}.
More recently, registries for pre-trained models, \eg Hugging Face, expose artifacts and informal model cards, but in an effort to keep up with rapid change they omit normalized dependency and capability schemas, hampering reuse~\cite{JiangPTMReuse,jiang2024peatmoss,Jiang2025Naming}.
Meanwhile, emerging agent protocols, including the Model Context Protocol (MCP)~\cite{mcp-registry-repo,mcp-registry-blog} and the Agent Name Service (ANS)~\cite{ans},
strengthen connectivity and naming but stop short of delivering a \textit{registry layer}.
The result is a fragmented landscape lacking features such as capability evidence and lifecycle status.

\begin{figure}[t]
  \centering
  \includegraphics[width=\columnwidth]{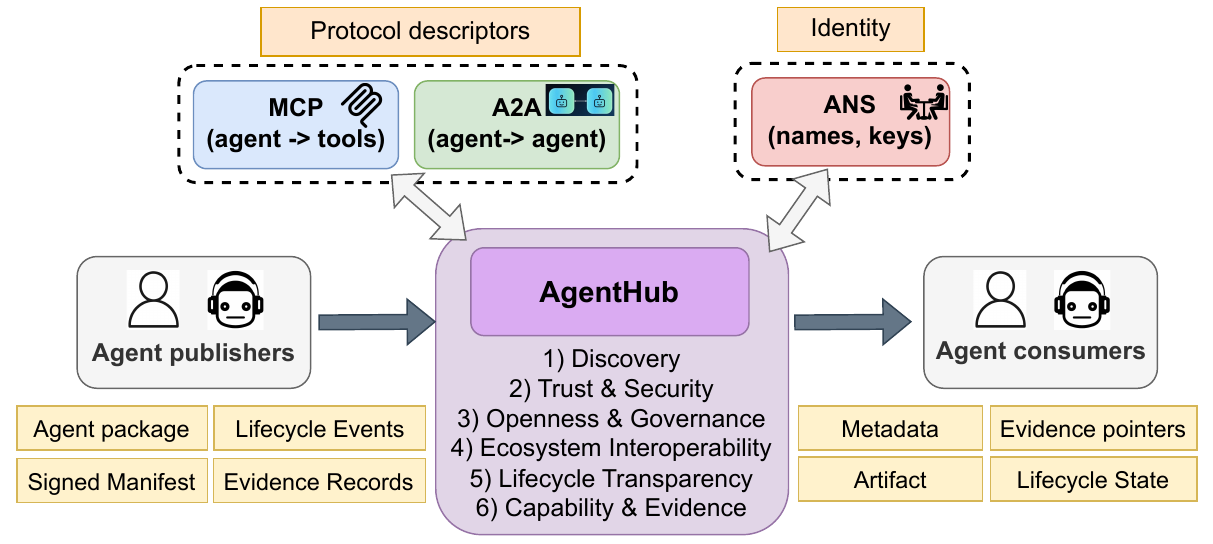}
  \caption{Conceptual view of \textit{AgentHub}, illustrating how publishers, identities, and agent protocols might interact.}
  \label{fig:agenthub-arch}
\end{figure}


We envision \emph{AgentHub} as a registry layer that sits between emerging agent protocols (\eg MCP, A2A, ANS) and the workflows that publish, discover, and reuse agents (\cref{fig:agenthub-arch}). 
AgentHub shares baseline registry concerns with earlier ecosystems, such as namespaces, versioning, integrity, and governance.
However, agents' autonomy, dynamic composition, continual evolution, and scale impose requirements that are difficult to treat as documentation problems.
Automatable reuse requires machine-checkable capability schemas and evidence (\cref{sec:Research-Capabilities}), and continual evolution requires lifecycle visibility and fast revocation (\cref{sec:Research-Lifecycles}).
Cross-protocol composition requires interoperability across descriptor dialects (\cref{sec:Research-Interop}).
Finally, scaling publishers and consumers beyond humans raises governance and security requirements that must be enforceable by default (\cref{sec:Research-OpennessAndGovernance,sec:Research-TrustAndSecurity}).

This paper contributes an agenda and a substrate for testing it.
We (i) distill lessons from mature software registries and emerging agent directory/protocol efforts into six requirements for agent-sharing infrastructure (\cref{fig:agenthub-research-agenda}); 
(ii) describe an open-source reference implementation that operationalizes these requirements via a canonical manifest, version-bound evidence records, and an append-only lifecycle log, with explicit interfaces for search, resolution, and verification (\cref{sec:Prototype}); 
and (iii) report initial discovery results and outline a broader evaluation program that targets each requirement, including deployability costs such as latency, storage growth, and scalability (\crefrange{sec:Eval}{sec:next-steps}).
Our aim is to make the design space for agent sharing infrastructure explicit, testable, and iteratively refinable by the community; we submit this work-in-progress to JAWS to solicit feedback on whether the six requirements and the prototype’s manifest/evidence/lifecycle contract capture the right minimal core for safe, automatable agent reuse.

\GKT{@Tanmay: The figure is looking nice but can you remove the "-" in front of the items in the central box? Can you also find a way to replace "..." with the remaining items from our agenda/vision?  I am now realizing I put this comment near the second figure. I meant to put near Figure 1.}

\begin{table*}[t]
  \caption{
  Aspects of existing software registries, and implications for AgentHub
  }
  \ifARXIV
  \else
  \small
  \fi
  \label{tab:registry-lessons-single}
  \centering
  \setlength{\tabcolsep}{4pt} 
  \renewcommand{\arraystretch}{1} 
  \begin{tabularx}{\textwidth}{@{}p{3cm}XX@{}}
    \toprule
    \textbf{Aspects Required} & \textbf{Examples in Software Registries} & \textbf{Implication For AgentHub} \\
    \midrule
    Metadata and dependency schema & 
    Package manifests (npm, PyPI, HF cards) encode machine-readable metadata and support versioned dependency graphs with auto-resolution (npm, Maven) \cite{npm-trusted,pypi-pep458,hf-modelcards} &
    Shared ontology (capabilities, I/O, protocols, provenance) with explicit agent--agent/service dependencies for reproducibility \\
    \addlinespace
    Trust and provenance & 
    Signing and provenance (npm ECDSA; PyPI TUF) \cite{npm-signatures,pypi-pep458} &
    Signed manifests and reproducibility attestations \\
    \addlinespace
    Governance and lifecycle management &
    Open vs.\ curated submission models (npm, CRAN/app stores) \cite{npm-2fa,cran-policies,cran-checklist} and update/revocation mechanisms (PyPI TUF) \cite{pypi-pep458} &
    Hybrid governance: open submissions with vetting for high-risk agents, plus explicit lifecycle states (active/deprecated/retired) and emergency removal paths \\
    \addlinespace
    Quality signals &
    Ratings, downloads, badges &
    Stats, ratings, benchmarks, audit badges for selection \\
    \bottomrule
  \end{tabularx}
\end{table*}

\section{Background and Related Work}
\label{sec:Background}

To support our vision, we analyze both
  pre-agent software registries
  and
  recent work on agent directory services.

\subsection{Lessons from Pre-Agent Software Registries}
\label{sec:lessons}

\JD{For readability, Force tabe 1 to appear on page 2 (move the declaraton updated in the tex)}
An agent registry can draw lessons from earlier registries (\cref{tab:registry-lessons-single}).
\WJ{This sentence is a bit odd. Should we say ``Table 1 summarized xxx''?}

\subsubsection{Metadata}
Programming language package registries such as npm, PyPI, Maven Central, and CRAN demonstrate the importance of structured metadata and explicit dependency graphs.
Package manifests (\eg \texttt{package.json}, \texttt{POM.xml}) encode versioning, licensing, and dependency constraints in machine-readable form, enabling automated resolution and reproducible builds~\cite{pypi-pep458,npm-trusted,hf-modelcards}.\WJ{Can/Should we also mention that AgentHub can provide `lockfile` to support better reproducible build?}
SBOM standards require explicit declaration of components and dependency relationships for provenance and traceability~\cite{ntia-sbom-minimum,ntia-sbom-framing}; emerging AI/ML BOMs extend the same idea to models, datasets, and configurations~\cite{xia2024trustsoftwaresupplychains, bennet2025SPDXAIBOM}.
Hugging Face relies primarily on model cards with limited dependency information~\cite{JiangPTMReuse}.
For example, while some cards reference required libraries, pretrained checkpoints, or paired datasets, these links are neither mandatory nor normalized into a dependency graph schema.
This lack of standardized schemas leads to inconsistent naming and hampers automated reproducibility~\cite{Jiang2025Naming, JiangScored22, jiang2024peatmoss}.
The lesson for AgentHub is that \textit{agent metadata must go beyond identifiers to include standardized schemas that capture capabilities, input--output modalities, protocol requirements, and declared dependencies}. 

\subsubsection{Provenance}
Trust and provenance mechanisms are central to registry design.
Maven Central requires every artifact to be signed with a PGP key; npm supports registry signatures and ``trusted publishing'' using OIDC~\cite{npm-trusted,npm-signatures}; and PyPI is adopting The Update Framework (TUF) 
for signed metadata~\cite{pypi-pep458}.
For agents, provenance is especially critical because dynamic tool bindings and environment access amplify the risks of impersonation or poisoning.
Accordingly, \textit{AgentHub should require signed metadata and reproducibility attestations for all entries}.

\subsubsection{Governance \& Submission Policies}
Governance models highlight trade-offs between openness and safety.
Npm and PyPI accept broad participation with light pre-checks, while CRAN and app stores such as Apple’s App Store impose strict manual review ~\cite{apple-appstore-guidelines, npm-2fa}.
Ecosystems also implement revocation: app stores can remotely disable malicious apps, and PyPI can yank faulty releases~\cite{pypi-pep458}.
Notably, the leftpad incident in npm illustrated the ecosystem-wide disruption that can follow from a poorly governed removal~\cite{npm-leftpad, theregister-leftpad}.
\textit{Governance requires care and is non-obvious with autonomous agents.}

\subsubsection{Discovery \& Quality Signals}
Registries, model hubs, and 
extension marketplaces provide user ratings, download counts, verification badges, and metadata-rich model cards~\cite{hf-modelcards,hf-popularity}.
These signals allow users to identify reputable contributions at scale.
\textit{AgentHub should adopt reputation systems such as usage statistics, audits, or benchmark results to complement technical metadata}.



\subsection{Related Work on Emerging Agentic Systems}
\label{sec:bg-agentInfra}

Several prior works have targeted a related use case: using agents for the services they provide.
However, these works have not considered the registry use case, where actors can go to identify agents and agent components.
The Agent Name Service (ANS) proposes a DNS-style directory for agents, offering secure, protocol-agnostic naming and discovery~\cite{ans}. 
The Agent Capability Negotiation and Binding Protocol (ACNBP) builds on ANS to enable secure capability negotiation among heterogeneous agents~\cite{acnbp}.
The NANDA Index introduces a decentralized, peer-to-peer index of agents with cryptographically verifiable ``AgentFacts'' attesting to capabilities and permissions~\cite{nanda}. 
Similarly, the MCP Registry catalogs MCP servers and their tools~\cite{mcp-registry-repo,mcp-registry-blog}, improving discoverability within the MCP ecosystem. 
Finally, curated marketplaces such as the ChatGPT Plugin Store or Alexa Skills Store show how policy-enforced ecosystems can scale with user trust~\cite{chatgpt-plugins, alexa-skill-ecosystem, alexa-skills-policy}.
Vendor SDKs are beginning to support an \textit{agent-make-agent} pattern; for example, Anthropic's Claude Agent can also generate orchestrated subagents~\cite{AnthropicAgentSDK,AnthropicSubagents}.

In formulating AgentHub, we observe that the emerging set of capabilities provided by prior works are necessary but not sufficient for the registry use case and the agent-make-agent scenario.
Addressing this demand requires new infrastructure, moving beyond technical protocols or metadata-only overlays to ensure transparency, interoperability, and accountability, which we outline subsequently in the research agenda (\cref{sec:Research-Agenda}).





\section{Research Agenda}
\label{sec:Research-Agenda}

We distinguish challenges
  shared by all registries (\cref{sec:agenda-common})
  from those unique to agents (\cref{sec:agenda-agent}).
We derived the AgentHub agenda as a scoped synthesis of two evidence streams: mature registry ecosystems for software and models, and emerging infrastructure for LLM-based agents.
First, we reviewed the design and operational practices of widely used package and artifact registries (\eg namespaces, versioning, provenance, governance, and security controls) alongside the software-engineering literature that analyzes their ecosystem dynamics and failure modes.
In parallel, we examined recent agent directory and protocol efforts (\eg agent cards and tool descriptors, agent-to-agent interaction protocols, and registry-like services) and extracted recurring friction points that arise when reuse becomes automated and cross-protocol.
We then iteratively grouped these observations into a small set of agent challenges and mapped each challenge to the registry mechanisms required to mitigate it.
The result is the set of six requirements in \cref{sec:agenda-agent}, organized in \cref{fig:agenthub-research-agenda} to make the traceability from requirements to challenges and objectives explicit.
This process is intended to be reproducible and extensible, so that as agent ecosystems mature, additional challenges can be incorporated and re-mapped without changing the core principle that registry mechanisms must be enforceable and machine-actionable to support autonomous reuse.
  
\ifARXIV
    \begin{figure*}[ht!]
\else
    \begin{figure}[ht!]
\fi

  \centering
\ifARXIV
  \includegraphics[width=0.9\linewidth]{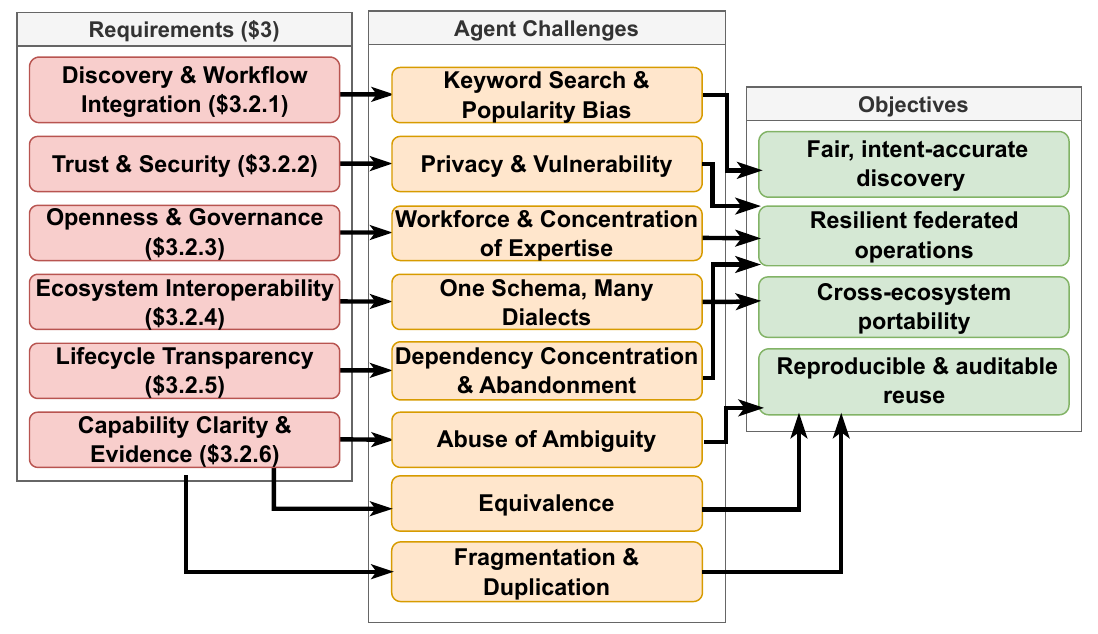}
\else
  \includegraphics[width=\columnwidth]{figures/Requirements_figure.pdf}
\fi
  \caption{
  Research agenda for AgentHub, showing how six requirements (§3.2) encounter specific challenges, motivating research directions toward objectives of reproducibility, portability, resilience, and fair discovery.
  \JD{Editing pass update: Align figure prose and order with the final content of \$3}
  }
  \label{fig:agenthub-research-agenda}
\ifARXIV
    \end{figure*}
\else
    \end{figure}
\fi

\GKT{I have been chatting with Erik about the confusing rendering and structure of headings in this section.
I think much of the problem stems from this subsection effectively being a section.
I think the Research Agenda section can stand on its own.
This section can be a discussion of Common Registry Challenges and each of the challenges would be a subsection.
This will result in a more effective rendering of the headings, which will have more pop than the underlined paragraph headings within each challenge.
I hope this makes sense. I know Jamie is fond of effective underlining. With this change, I think we can preserve the nice emphasis on our paragraphs while having more clarity about the challenges. subsubsection doesn't have much pop in this template, so when we use it, it should probably be the deepest level of heading we use.}
\EP{Just changed it, George. Let me know your thoughts.}
\subsection{Common Registry Challenges}
\label{sec:agenda-common}

As indicated in~\cref{tab:registry-lessons-single},
lessons learned from mature registries set the baseline:
  entries must carry structured manifests and dependency graphs for reproducibility;
  publishers and artifacts must be authenticated with signed metadata and public transparency logs;
  namespaces and lifecycle actions (publish, deprecate, revoke) must be governed;
  and
  the system must remain open and usable through simple APIs.
Proven machinery such as PURL-style identifiers and existing code and model hubs should be incorporated through integration and adaptation, not re-invention.
Since nearly all mainstream registries are centralized and host both metadata and artifacts, we expect similar centralization to be valuable for agent-related artifact management.
However, running agents requires substantial hardware, so directory services will likely be necessary for that use case (\cref{sec:bg-agentInfra}).

Next we present our research agenda in~\cref{fig:agenthub-research-agenda}:
  \textit{What changes when a registry's artifacts, and some of its actors, are Agents?}

\subsection{Agent-Specific Considerations}
\label{sec:agenda-agent}

\subsubsection{{Discovery and Workflow Integration}}
\label{sec:Research-Discoverability}

In an agent ecosystem, the ``users'' of the registry are both developers and agents that act autonomously at scale, so discovery must be programmatic, rank by capability-and-evidence fit rather than popularity, and integrate directly into planning, CI/CD, and orchestration loops.


\emph{Challenge: Keyword Search \& Popularity Bias.}
Agents that recommend or install one another can create feedback loops that instill mediocre or unsafe entries.
Experience from software registries highlights how naive signals mislead: Zerouali \etal show that different popularity metrics in npm yield inconsistent results~\cite{Zerouali2019}, underlining how reliance on downloads or stars can mislead users; and Jones \etal observed that model popularity on HuggingFace correlates strongly with documentation quality~\cite{jones2024knowhuggingfacesystematic}, suggesting that discovery in AgentHub would similarly benefit from incentives for high-quality documentation.
Agents benefit greatly if those signals are captured as structured metadata and verifiable evidence.
Without workflow integration, agents may fall back to ad-hoc search, undermining adoption and reproducibility. 

\emph{Solution:}
AgentHub will improve current practice by combining keyword search with structured metadata and evidence linked to benchmarks, incorporating lifecycle states to avoid unsafe or outdated agents. Evaluation should identify which signals such as metadata and documentation quality ensure reliable discovery. Discovery should match user goals to agents by capabilities and evidence, ensuring that popularity does not overshadow quality. This resembles work on recommendation systems for software libraries like Code Librarian, which uses contextual signals to suggest packages~\cite{tao2023codelibrarian}. Research benchmarking discovery is necessary to measure precision, recall, and resilience under ecosystem changes.


\subsubsection{{Trust and Security}}
\label{sec:Research-TrustAndSecurity}
Autonomous composition widens the attack surface: agents can install, call, or even generate other agents, so identity, provenance, and revocation must be machine-enforceable.
Entries should use signed manifests, verified namespace control, and provenance for builds, models, and datasets.
A useful precedent comes from the supply-chain domain: the SLSA v1.1 Verification Summary Attestation (VSA)~\cite{slsa-vsa} standardizes how to publish structured, signed evidence of checks performed on artifacts, and AgentHub could adopt an analogous mechanism to keep validation auditable and portable.
Such evidence must be authenticated (with optional third-party attestations for sensitive cases), and revocation or key rotation must propagate quickly across mirrors.

\emph{Challenge: Privacy \& Vulnerability.}
In software registries, account hijacks and malicious updates already erode trust~\cite{Zimmermann2019}.
LLM-based agents add privacy-specific attack surfaces: misconfiguration or weak governance can leak sensitive data or intellectual property~\cite{wang-etal-2025-unveiling-privacy,zharmagambetov2025agentdamprivacyleakageevaluation}; training-data exposure can reconstruct confidential content~\cite{carlini2021extractingtrainingdatalarge}; and real incidents have leaked corporate data~\cite{samsung_chatgpt_forbes_2023}.
Multi-agent prompt injection can propagate malicious instructions and compromise collective decision-making~\cite{lee2024promptinfectionllmtollmprompt,shi2025promptinjectionattacktool}.
Privilege escalation, guardrail bypasses, weak output validation, and insufficient access control enable unauthorized tool use and code execution~\cite{narajala2025securinggenaimultiagentsystems}.
The attack frontier remains underexplored.

\emph{Solution:}
AgentHub will deploy defenses such as signed manifests, provenance attestations~\cite{pypi-pep458,npm-trusted}, verified namespaces via ANS~\cite{ans}, and protections against typosquatting and package-confu\-sion attacks~\cite{Zimmermann2019,jiang2025confuguardusingmetadatadetect}.
AI-specific risks include deserialization exploits and prompt-injection.
It is crucial to test whether defenses from human-mediated registries hold for autonomous agent systems.
New threats call for zero-trust privilege separation, runtime checks for I/O behavior, and privacy-preserving audit pipelines to verify evidence without leaks~\cite{he2025redteamingllmmultiagentsystems, narajala2025securinggenaimultiagentsystems, samsung_chatgpt_forbes_2023, wang-etal-2025-unveiling-privacy}.



\subsubsection{{Openness \& Governance}}
\label{sec:Research-OpennessAndGovernance}
\HP{Behnaz: talk about OpenSSF’s package deletion policies here}
\TS{Lets talk about it here instead of BG then}
At agent scale, publishers include both humans and automated pipelines, so governance must keep namespaces open and verifiable while preventing automated spam, squatting, and opaque takedowns.
Traditional software registries such as PyPI and npm succeeded not only by providing distribution infrastructure but by embracing openness: anyone could publish, namespaces were transparent, and governance processes were clear.
For AgentHub, these properties are even more critical.
Without mechanisms for open contribution, transparent review, and consistent namespace management, registries risk becoming closed silos controlled by a few vendors as ecosystems evolve.

\emph{Challenge: Workforce \& Concentration of Expertise.}
Many projects are maintained by a single person, creating bus-factor risk. 
Zimmermann \etal emphasize that the issue is less maintainer shortage than concentrated control~\cite{Zimmermann2019}.
Governance models differ: npm allows instant publication, while curated ecosystems like CRAN impose stricter checks.

Even if agents can take over some maintenance tasks, expertise may still concentrate in a few organizations or key models.
This creates the risk that critical agents depend on too few people.
To avoid such single points of failure, AgentHub governance must balance openness with safeguards: partial vetting for high-impact agents, incentives for broader participation, and clear processes to revoke unsafe agents.

\emph{Solution:}
Future work should investigate governance structures that combine openness with verifiability.
Possible directions include decentralized namespace assignment rooted in ANS, community-driven policy boards for resolving disputes, and auditable logs of publication and revocation decisions.
Comparative studies of centralized versus federated governance models can illuminate trade-offs in consistency, adoption, and resilience.
Open questions include how to embed transparency in decision-making without sacrificing agility, and how to make governance mechanisms both fair across domains and enforceable at internet scale.

\subsubsection{{Ecosystem Interoperability}}
\label{sec:Research-Interop}

Planners and orchestrators compose agents and tools across protocols; cross-protocol operation requires a shared metadata core with protocol-specific extensions so intent-based queries can compare agents without losing meaning.
It also requires portable, signed manifests and SBOM-style dependency graphs so tools, models, datasets, and services remain traceable across registries (\eg npm, PyPI, model hubs) and agent protocols.
Many model hubs still rely on ad-hoc files (\eg free-form \texttt{config.json}), hurting portability and automated reuse.
\WJ{We should echo the requirements for better manifest files and SBOMs. These two are the key parts to support ecosystem interoperability in the traditional SW supply chains.}
\EP{addressed Wenxin's comment}

\emph{Challenge: One Schema, Many Dialects.}
Descriptors in MCP (tools), A2A (roles/behaviors), and ACP (message schemas/policies) use different primitives and evolve at different speeds.
Without a standardized, signed manifest, semantics may be lost in translation, caches drift across mirrors, and naive popularity metrics dominate cross-protocol rankings while ignoring evidence quality and freshness.
Stable identifiers for referenced artifacts are also missing, making cross-registry links fragile.



\emph{Solution:}
We propose a compact capability ontology and a canonical manifest with required fields (capabilities, I/O modalities, protocol bindings, permissions, SBOM-style dependencies) and optional per-ecosystem extensions.
Declarative adapters can then map descriptors to this core and back, validated via round-trip conformance tests.
Stable cross-registry identifiers (\eg Purl-style) should be introduced and tested end-to-end with npm, PyPI, and model hubs.
Early benchmarks could measure cross-protocol discovery in terms of precision/recall for intent queries, preservation of evidence link, freshness under churn, and ranking fairness as ecosystems scale.


\subsubsection{{Lifecycle Transparency}}
\label{sec:Research-Lifecycles}

Software registries expose version history and revocations; autonomous agents need richer lifecycle states--active, deprecated, rotated, retired, or revoked--with timestamps and rationales.
Because agents evolve dynamically and can continue acting without human oversight, lifecycle visibility is essential for safe reuse and governance.
Discovery should respect these states so outdated entries do not appear healthy by default, and mirrors should propagate state changes within bounded freshness windows.




\emph{Challenge: Dependency Concentration \& Abandonment.}
The leftpad incident showed how removing even trivial packages disrupted thousands of projects.
Zimmermann \etal found that a small number of npm maintainer accounts control most packages~\cite{Zimmermann2019}, while Zerouali \etal note that popularity often masks inactivity~\cite{Zerouali2019}.
Agents face parallel risks: if many depend on a single base agent or are generated by one entity, failure or compromise could cascade broadly.
Abandonment may occur not only when humans leave but also when autonomous agents stop updating, leaving stale yet discoverable entries in circulation.


\emph{Solution:}
Future work should define lifecycle metadata standards with clear states, timestamps, and rationales, along with monitoring to detect abandonment or unexpected behavioral drift.
Research must also address how deprecated or revoked agents should appear in discovery, who has authority to mark or revoke them, and how federated mirrors should coordinate state changes.
Comparing agent lifecycles with traditional software lifecycles may reveal where new loops--such as agents participating in design and implementation--demand stronger provenance, transparency, and control.

\subsubsection{{Capability Clarity \& Evidence}}
\label{sec:Research-Capabilities}
\WJ{I feel like this subsection belongs to the concept of ``AI governance'' (\$3.4). Should we merge these two sections?}\WJ{There are also a lot of government documents calling for better AI governance that we could cite. For example, \url{https://cdn.governance.ai/GovAI-Research-Agenda.pdf} and \url{https://www.whitehouse.gov/wp-content/uploads/2025/07/Americas-AI-Action-Plan.pdf}}
\JD{I do not think that capability tracking is quite the same thing as governance (which I think of more as thinking about what capabilities *should* be possible, rather than tracking what currently *is* possible).}
\WJ{I agree that capability tracking is more about what *should* be possible. I wonder if that is still under the broader concept of ``governance''}

Registries for conventional software rely on manifests to make artifacts understandable and reproducible, but agents need a richer contract.
For autonomous agents that compose tools and other agents at runtime, manifests must express runtime permissions, preconditions, environment bindings, and protocol roles. 
A useful analogy is Android’s permission model~\cite{android_permissions}, where apps declare capabilities in a manifest that the OS validates during installation and use.
Similarly, AgentHub manifests could expose agent capabilities in machine-readable form, enabling tools to flag over-privileged or under-evidenced agents before adoption.
Engineers and agents should be able to plan for behaviors, not just observe them. 


\emph{Challenge: Fragmentation \& Duplication. }
As npm's \linebreak ``micro-packages'' created brittle dependency chains~\cite{kula2017impactmicropackagesempiricalstudy,Abdalkareem2017TrivialPackages}, agent ecosystems may spawn ``micro-agents'' with overlapping functions.
Because agents can autonomously query registries and compose others, duplication and fragility can spread dynamically, enlarging the attack surface and degrading reliability, especially at scale as agents evolve and reimplement overlapping capabilities.

\emph{Challenge: Equivalence.}
Beyond functional duplication, a deeper challenge is determining when two agents are truly equivalent.
At a syntactic level, the same agent may appear across multiple repositories, creating confusion about which copy is authoritative. 
At a semantic level, multiple agents may claim the same capabilities but diverge in behavior due to nondeterminism and evolution in models, changes in context, and even hardware selection.
Addressing this requires more than metadata alignment: AgentHub should support persistent identifiers, cross-registry attestations, and re-executable evidence pipelines to assess whether two agents are truly equivalent.


\emph{Challenge: Abuse of Ambiguity.}
Ambiguity and missing metadata enable attacks in software registries~\cite{JiangPTMReuse}; agents face the same risk.
Unclear manifests let adversaries mimic popular entries or collaborators, echoing typosquatting and account hijackings in npm~\cite{Zimmermann2019}.
Agents' autonomous installation/generation can accelerate such propagation unless strong signing and evidence requirements are enforced.
%

%
\emph{Solution:}
AgentHub will
  (i) standardize machine-read\-able capability schemas covering capabilities, modalities, protocol roles, and dependencies enforced at publish time~\cite{JiangPTMReuse,Jiang2025Naming};
  (ii) support lightweight, re-executable (idempotent) evidence pipelines linking claims to traces or benchmark runs across versions;
  and
  (iii) add badges or similar checks as digital nudges~\cite{brown2019nudges}. 

\section{Prototype}
\label{sec:Prototype}
AgentHub is grounded in a concrete, open-source reference implementation that makes the agenda's mechanisms directly testable. 
The prototype is intentionally not an agent runtime. 
Instead, it is a registry layer that supports publishing, discovery, and reuse of agents under machine-checkable contracts, such as standardized capability manifests and evidence, lifecycle transparency, protocol interoperability, and a baseline trust and security model suitable for automated reuse.
The goal of the reference implementation is to provide a minimal but complete substrate that can publish, index, search, resolve, verify, and evolve so that the requirements in \cref{fig:agenthub-research-agenda} can be evaluated empirically and iterated with the community.

\subsection{Architecture \& Scope}
The prototype follows a design that stresses mutable metadata and immutable artifacts.
Agent entries are represented as versioned snapshots of a canonical manifest stored immutably by content hash, while the registry maintains mutable metadata (indexes, lifecycle state, and evidence summaries) that can evolve without changing the underlying artifact.
This makes verification straightforward, so clients can recompute the hash of a retrieved manifest and compare it to the registry's recorded digest, and signatures can be defined over that digest rather than over an ambiguous, mutable document.

\JD{Suggest we use a bullet list here}
At a high level, the system exposes a small set of operations that map directly to the user and agent workflows described in \cref{sec:Research-Discoverability}: (i) publish an agent version, (ii) search and rank candidates, (iii) resolve a stable identifier and version range to a concrete version and endpoint bindings, (iv) verify provenance signals, (v) attach evidence records, and (vi) append lifecycle events that affect discovery and resolution.
Discovery is treated as a first-class interface, because in an agent ecosystem the users of the registry include both humans and agents, and discovery must integrate into planning and orchestration loops rather than requiring manual browsing.

\subsection{Canonical Manifest, Evidence Records, and Lifecycle Events}
A central deliverable of the prototype is a core manifest with publish-time validation. 
The manifest provides the minimum required fields that enable automation and reproducibility across heterogeneous ecosystems, including stable identifiers; declared capabilities and I/O modalities; protocol bindings; runtime permissions and preconditions; SBOM-style dependency references; explicit lifecycle state; and an evidence policy that declares what kinds of evidence are expected and how freshness is interpreted.
Optional per-ecosystem extensions allow native descriptors to be carried without forcing premature convergence.
This design follows our "one schema, many dialects" requirement, as we will keep a required core that is stable enough to be validated and indexed, while preserving protocol-specific descriptors in an extension space (\cref{fig:agenthub-manifest}).

\ifARXIV
    \begin{figure*}[ht!]
\else
    \begin{figure}[ht!]
\fi

  \centering
\ifARXIV
  \includegraphics[width=0.9\linewidth]{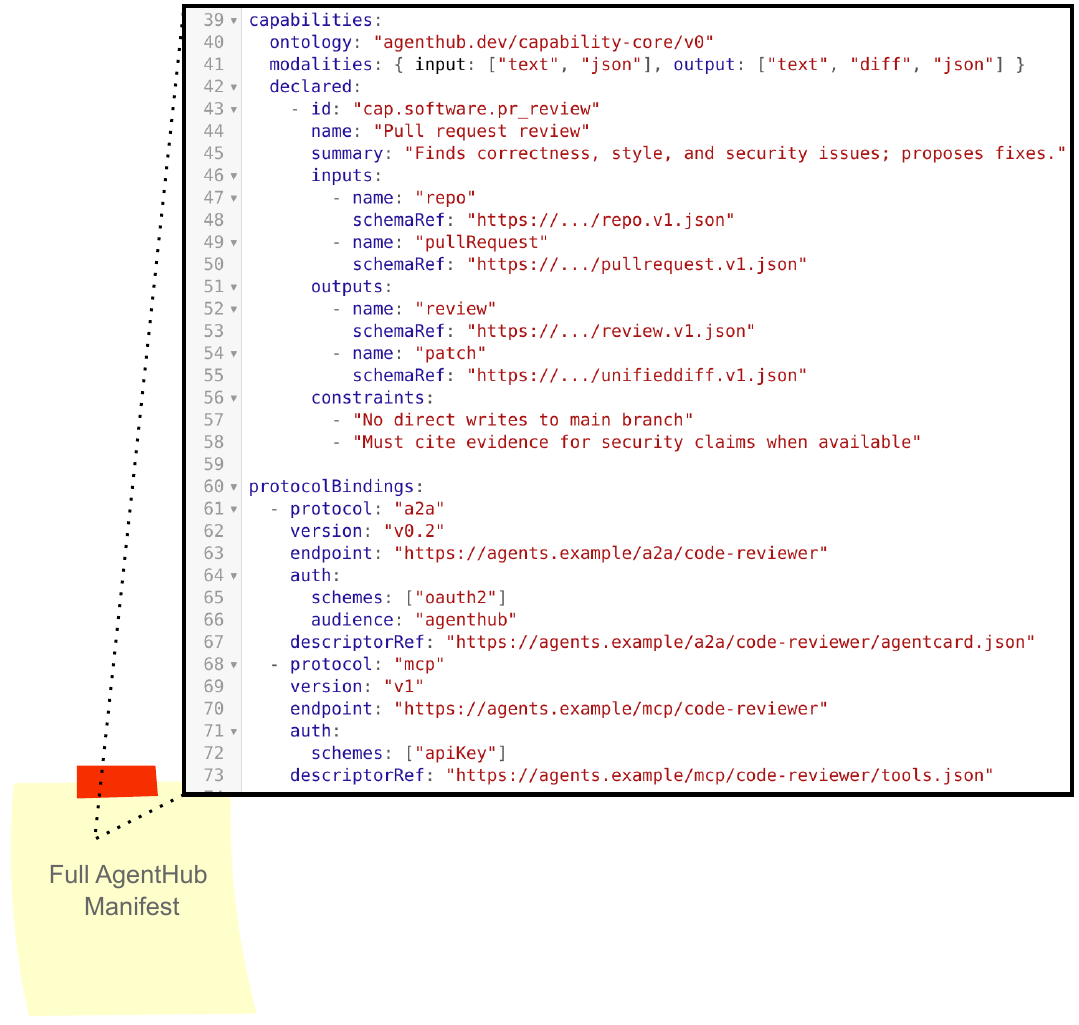}
\else
  \includegraphics[width=\columnwidth]{figures/Manifest.pdf}
\fi
  \caption{
  Zoomed excerpt of an example AgentHub manifest illustrating capability declarations (I/O schemas and constraints) and protocol bindings for multiple ecosystems (A2A, MCP). AgentHub validates and indexes these fields to support intent-accurate discovery and cross-protocol resolution.
  }
  \label{fig:agenthub-manifest}
\ifARXIV
    \end{figure*}
\else
    \end{figure}
\fi

At publish time, the core manifest is validated under a strict schema, canonicalized deterministically, and stored as an immutable blob keyed by its hash.
The schema is designed to be machine-actionable with capabilities including stable identifiers and, where possible, structured input/output expectations; protocol bindings name the interaction dialect (e.g., MCP vs. A2A) and provide resolvable endpoints; permissions declare resource access explicitly (network, filesystem, tool access, data handling), so that automated selection can incorporate risk constraints rather than relying on prose.

The prototype also defines an evidence record as a separate, version-bound object that links a capability claim (or a broader "fitness claim") to auditable artifacts. Each record binds to a specific agent version and manifest digest, and carries (i) the method/recipe used to produce the evidence (e.g., harness version, container digest/workflow reference), (ii) inputs (config, seeds, environment), (iii) outputs as immutable artifacts with hashes, and (iv) optional attestations.
Evidence is stored immutably and referenced from the registry, enabling third parties to reproduce or independently re-run checks.
This operationalizes our argument that agents benefit when discovery signals are captured as structured metadata and verifiable evidence rather than popularity or documentation alone.

Finally, the prototype treats lifecycle as an append-only event log rather than a single mutable flag.
Agent versions move through explicit states (e.g., active, deprecated, retired, revoked), and each state transition is recorded with a timestamp and rationale. The registry's default behaviors respect these states. 
Resolution must not return revoked/retired versions, and discovery must incorporate lifecycle state into ranking and filtering. 
This is essential for lifecycle visibility and fast revocation in ecosystems where reuse is automated and version churn is frequent.

\subsection{Interoperability and Discovery as Registry Mechanisms}
Interoperability in our prototype will be implemented as declaractive adapters that map native protocol descriptors into the core manifest and back. MCP and A2A use different primitives and evolve at different speeds, so a standardized, signed manifest is needed to prevent semantic loss and cache drift across mirrors. 
In the prototype, protocol bindings provide a stable, comparable surface (capabilities, modalities, endpoints, auth requirements), while the original descriptor (e.g., an A2A Agent Card or MCP tool descriptor) can be stored in \texttt{extensions} for lossless round-trip translation where possible.
The concrete outcome is testable: adapters are validated through round-trip conformance tests that check preservation of core semantics and the stability of cross-registry identifiers.

Discovery is implemented as a registry concern because it is one of the primary ways automation will consume AgentHub.
Our prototype will provide programmatic search over manifest fields and evidence signals, and a resolution interface that maps stable identifiers and version ranges to an exact version and protocol endpoints.
Ranking is designed to be evidence forward.
Rather than defaulting to popularity proxies, the registry returns candidates with explicit reasons such as relevance from structured metadata, evidence coverage and freshness, lifecycle state, and compatibility constraints. This directly responds to the "keyword search and popularity bias" challenge identified in \cref{fig:agenthub-research-agenda}.

\section{Preliminary Evaluation}
\label{sec:Eval}
AgentHub's research agenda (\cref{fig:agenthub-research-agenda}) frames six requirements that collectively target four objectives: fair, intent-accurate discovery; resilient federated operations; cross-ecosystem portability; and reproducible, auditable reuse.
\JD{Don't explain the future work here. Just explain what we did for the preliminary evaluation. These opening remarks should explain what we are trying to measure and why.}
In addition to early evaluations on discovery (discussed below), we propose a concrete evaluation program that combines (i) controlled benchmarks that isolate mechanisms and quantify trade-offs, (ii) adversarial experiments that test whether defenses remain effective under autonomous reuse, and (iii) longitudinal, deployment-level measurements that capture governance and ecosystem dynamics. These evaluation plans are discussed below.

\JD{Need subsections for METHOD and RESULTS}

\subsection{Evaluating Discovery and Workflow Integration}
\subsubsection{Method:}
We implemented a two-stage recommendation pipeline on top of ANS that recommends A2A agents using their Agent Cards as structured, comparable descriptions. \cref{fig:agenthub-discovery-pipeline} summarizes the interaction pattern, where a client submits a query, the recommendation system consults the registry to retrieve and rank candidates, and the selected agent is returned to the client. 

\ifARXIV
    \begin{figure*}[ht!]
\else
    \begin{figure}[ht!]
\fi

  \centering
\ifARXIV
  \includegraphics[width=0.9\linewidth]{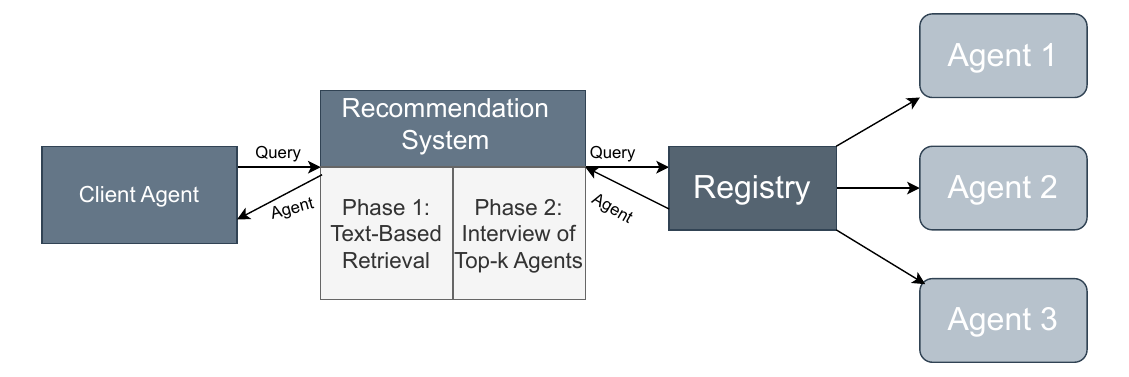}
\else
  \includegraphics[width=\columnwidth]{figures/agentworkflow.drawio.pdf}
\fi
  \caption{
  Two-stage discovery workflow used in our initial evaluation. 
  A client query is handled by a recommendation system that consults the registry for candidate agents, then returns a ranked selection to the client.
  }
  \label{fig:agenthub-discovery-pipeline}
\ifARXIV
    \end{figure*}
\else
    \end{figure}
\fi

In stage one, the recommendation system applies text-based retrieval over A2A agent cards to identify candidate agents relevant to a user query. 
We evaluate three retrieval strategies: a lexical approach (BM25), a semantic embedding-based approach, and a hybrid method that combines lexical and semantic signals.  \cite{8442792, bm25s, sultania2024domainspecificquestionansweringhybrid}
\JD{Cite these methods}
This stage is designed to efficiently narrow the candidate set using static, declarative agent descriptions.

In stage two, the highest-ranked candidates from stage one are further evaluated using a structured, interview-based verification process.
This phase engages agents directly to elicit behavioral evidence, allowing the system to assess whether an agent’s actual capabilities align with the inferred intent of the query. 
Importantly, this stage is intended to produce explicit evidence of suitability, rather than re-rank candidates based on textual similarity.

\JD{More detail is needed on this part. What is `live server', why `four pairs', give an example `natural-language query', and how did we decide `designated as the best match'?}

The evaluation registers eight A2A-based agents as live servers, meaning each agent is deployed as an independently running service that exposes its A2A interface and responds to real-time queries. The agents are intentionally organized into four pairs with overlapping but distinct capabilities to create controlled ambiguity during discovery (e.g., multiple debugging or performance-related agents with different specializations). We evaluate the system using 24 semi-ambiguous natural-language queries, each designed so that more than one agent could plausibly address the request, but only one agent represents the intended primary capability.
For example, the query “Agent that diagnoses backend crashes occurring only under concurrent load” could be handled by several debugging-related agents, but the “code-debugging-assistant” is designated as the best match because it explicitly targets concurrency-related logic faults. Ground-truth labels for all queries were assigned based on domain expertise and each agent’s intended functional scope. This dataset design reflects realistic discovery scenarios in which advertised capabilities overlap, requiring deeper behavioral validation beyond surface-level descriptions.

The study measures ranking quality and efficiency using Precision@1, Recall@3, and latency, averaged across multiple runs to account for LLM nondeterminism in the interview stage.  Precision@1 measures the fraction of queries for which the top-ranked (recommended) agent is the correct one. 
Recall@3 measures the fraction of queries for which the correct agent appears within the top three results. 
Latency captures the total end-to-end time required to process a query and return a recommendation.
We first compare retrieval-only performance across lexical, semantic, and hybrid methods, and then evaluate the full two-phase pipeline that supplements retrieval with structured interview-based verification.  


\subsubsection{Results:}
Hybrid retrieval achieves the best balance among retrieval-only methods (Precision@1 83.33\%, Recall@3 95.83\%, latency 0.57s), while the full pipeline maintains high Recall@3 (98.61\% $\pm$ 1.97\%) at substantially higher latency (101.30s $\pm$ 7.25s), explicitly positioning the interview stage as an evidence-producing mechanism rather than pure ranking booster. 
\cref{tab:combined_results} summarizes the results of different phases and retrieval methods.

\begin{table}[t]
\centering
\caption{Phase~1 Retrieval vs. Two-Phase Pipeline}
\label{tab:combined_results}
\small
\begin{tabular}{lccc}
\hline
\textbf{Method} & \textbf{P@1} & \textbf{R@3} & \textbf{Lat (s)} \\
\hline
\multicolumn{4}{l}{\textit{Phase~1: Text Retrieval}} \\
BM25      & 75.00 & 91.67 & 0.01 \\
Semantic  & 66.67 & 100.00 & 0.85 \\
Hybrid    & 83.33 & 95.83 & 0.57 \\
\hline
\multicolumn{4}{l}{\textit{Phase~2: Hybrid + Interviews}} \\
Full Pipeline & 73.61 ± 1.97 & 98.61 ± 1.97 & 101.30 ± 7.25 \\
\hline
\end{tabular}
\vspace{3pt}
\end{table}

\section{Next Steps and Roadmap}
\label{sec:next-steps}

\JD{Right now we are mixing completed measurements, with (next part) conjecture. To avoid that, let's place the rest of this section in a section titled ``Next steps''. That will require some rework of section 6 (move the roadmap part back here).}
\EP{ACK}

We treat the results from \cref{sec:Eval} as an initial benchmark for the "Discovery \& Workflow Integration" requirement, and we will expand it in three ways. 
First, we will scale beyond a small, curated pool into a seed corpus that includes heterogeneous agent types (tool-using agents, agent-to-agent coordinators, domain-specialized micro agents) and deliberately introduce documentation and metadata variance, reflecting real registry messiness rather than uniform A2A-only artifacts.
Second, we will evaluate discovery not only by ranking metrics but also by downstream task success in workflows (e.g., whether a selected agent completes a CI/CD task under policy constraints), measuring end-to-end correctness and the cost of mis-selection.
Third, we will evaluate robustness and fairness by injecting popularity signals and churn, testing whether intent and evidence can dominate naive "stars/downloads" effects as ecosystems scale, consistent with the paper's warning that popularity loops can dominate discovery.

\subsection{Evaluating Trust and Security}
AgentHub argues that autonomous composition widens the attack surface and makes provenance, revocation, and enforceable identity central. 
The key evaluation question is not whether individual defenses exist, but whether they hold when (a) agents can publish agents, (b) agents can select agents, and (c) reuse is automated at scale.
We therefore will evaluate Trust \& Security through an explicit registry red-team suite that attempts to publish and propagate adversarial artifacts, including typosquatting and package-confusion analogs, identity spoofing, over-privileged manifests, malicious updates, and prompt-injection or tool-misuse behaviors embedded in agent workflows. 

Evaluation outcomes will include attack success rates under different enforcement policies (e.g., publish-time schema enforcement, signature requirements, namespace verification, evidence-gated promotion), time-to-detection and time-to-containment, and the degree to which security signals can be made machine-actionable without turning governance into opaque central control.
We will also explicitly test for agents that misrepresent capabilities for selection advantage, as this was an observed weakness during our discovery evaluation. We will introduce adversarial capability inflation agents and measure whether evidence requirements and behavioral checks can detect and penalize strategic misrepresentation.

\subsection{Evaluating Openness and Governance}
The paper motivates governance as a balance between openness (low-friction publishing, transparent namespaces) with safeguards against spam, squatting, and opaque takedowns.
We will therefore structure evaluation for this requirement in two phases. 
In the prototype phase, we run controlled publishing studies that measure submission friction, correctness of manifests under enforcement, and how often policy checks block legitimate submissions versus adversarial or low-quality ones.
In the deployment phase, we will instrument the registry for longitudinal metrics such as submission volume, acceptance/rejection rates by reason code, median time-to-resolution for disputes, rates of automated publishing versus human publishing, and the concentration of publishing authority across namespaces.

We will also run governance stress tests that simulate realistic failure modes such as coordinated spam attempts, namespace disputes, and emergency revocations. 
The goal is to empirically characterize the trade-offs between curated and open models for agents, under the distinct condition that many publishers may themselves be automated pipelines.

\subsection{Evaluating Ecosystem Interoperability}
In AgentHub, we see interoperability as more than just supporting MCP and A2A; it is specifically "one schema, many dialects," with canonical manifests and declarative adapters that will be validated by round-trip conformance tests so that semantics are preserved across protocol translations. 
This requirement is well-suited to a rigorous, testable evaluation in which we will define a suite of reference agents expressed natively in different ecosystems (e.g., MCP tool descriptors and A2A agent cards), translate them into the AgentHub core manifest, and translate back, measuring round-trip fidelity at the level of capabilities, modalities, permissions, dependencies, and evidence links.
We will complement this with cross-protocol discovery benchmarks, such as identical intent queries that should retrieve comparable agents regardless of whether the underlying descriptor dialect is MCP- or A2A-native. 
Evidence links should also remain resolvable and correctly bound to versions across protocol boundaries.

Interoperability evaluation will also include failure characterization to investigate which fields are lossy under translation, which protocol features cannot be expressed in a shared core without distortion, and what minimal portable contract is sufficient for safe cross-ecosystem reuse.

\subsection{Evaluating Lifecycle Transparency}
Lifecycle transparency becomes meaningful only if lifecycle state changes propagate quickly enough to affect automated reuse.
This paper argues for explicit states (active, deprecated, rotated, retired, revoked) with timestamps and rationales, and for coordination across federated mirrors.
We will evaluate this requirement via churn experiments that simulate realistic ecosystem dynamics including frequent version releases, dependency updates, emergency revocations, and key rotations.
We will measure propagation latency of lifecycle events across mirrors and caches, the degree to which clients honor lifecycle state in discovery and selection, and failure cases where stale mirrors lead to unsafe reuse.

\subsection{Evaluating Capability Clarity and Evidence}
Capability clarity and evidence is the requirement that most sharply differentiates AgentHub from existing directories. 
Our near-term priorities include enforcing machine-readable capability schemas at publish time and supporting lightweight, re-executable evidence pipelines that link capability claims to benchmark runs or traces across versions.
We will evaluate this along two axes: whether manifests become sufficiently precise to support automated planning and risk checks, and whether evidence is both (a) discriminative for selection and (b) reproducible and auditable.

On precision, we plan to test whether an enforced manifest schema reduces ambiguity and capability inflation by measuring inter-rater agreement among humans and agents when mapping natural-language intents to manifest claims, and by measuring how often over-privileged or under-specified agents can pass publish-time validations.
On evidence, we will treat evidence records as structured, signed, re-runnable artifacts and quantify (i) reproducibility across environments and repeated runs, (ii) stability under model or dependency updates, and (iii) the relationship between evidence signals and downstream task success.
Interview-based verification can be used as one evidence mechanism for ambiguous intents, and we will assess its value in terms of confidence/correctness trade-offs.


\JD{Consider a subsection titled `Practical consideratoins' where we write about, and measure, things like latency, storage cost, scalability}
\EP{We're squeezed on real estate.. shorten other eval sections to make room for practical considertions?}

\subsection{Practical Considerations}
Beyond ranking quality, we will report operational costs that determine deployability.
We will measure (i) publish-time overhead for validation, canonicalization, and signature checks, (ii) end-to-end query latency for search, resolution, and verification, and (iii) storage growth for manifests, evidence artifacts, and lifecycle logs. 
We will scale the number of registered agents and update churn to report throughput (publishes, queries), tail latency, and the dominant bottlenecks.

\subsection{Roadmap (Jan--Sep 2026).}
We treat this workshop paper as Phase 1 of a longer effort that culminates in a journal-length contribution and a reusable artifact suite.
Over the next eight months we will stabilize the core contract, strengthen interoperability and evidence plumbing, expand evaluation coverage across all six requirements, and release artifacts that others can adopt and extend.
\textbf{Jan--Mar:} freeze the v0.x manifest and evidence schemas; implement publish-time validation and canonicalization; add lifecycle events and signature verification; release a minimal CLI/API.
\textbf{Apr--May:} implement declarative adapters for at least two dialects (MCP and A2A) with round-trip conformance tests; evaluate lifecycle freshness and revocation propagation under churn and caching.
\textbf{Jun--Aug:} scale the seed corpus and intent suite; broaden from discovery metrics to workflow-level outcomes under policy constraints; add a robustness suite (inflation, malicious updates, typosquatting, prompt/tool misuse). 
\textbf{Sep:} package schemas, validators, adapters, benchmarks, and harnesses as a reproducible bundle and submit the journal-length manuscript.



\section{CONCLUSION}
\label{sec:Conclusion}
Agent ecosystems are scaling faster than the infrastructure needed to make them discoverable, comparable, and governable under automated reuse.
We advance AgentHub as a registry substrate and research agenda grounded in a reference implementation centered on (i) a canonical, publish-time validated manifest that is indexable yet extensible, (ii) version-bound evidence records that connect capability and fitness claims to auditable artifacts, and (iii) lifecycle states enforced by default in resolution and discovery.
Our discovery benchmark quantifies the value of structured, machine-actionable metadata for retrieval and clarifies when behavioral evidence is needed under ambiguous intents, motivating the broader evaluation program in \cref{sec:Eval}.
Together, AgentHub moves selection and governance signals out of prose and into enforceable, machine-checkable contracts, enabling auditable, comparable, and automatable agent reuse.

\textbf{Data Availability.} The artifact, including \texttt{manifest.yaml} and \texttt{evidence\_record.yaml}, is available at \cite{agenthub-artifact}.

\ifARXIV
\section*{Acknowledgments}

Davis acknowledges support from NSF awards \#2343596, \#2537308, and \#2452533.
Thiruvathukal and Läufer acknowledge support from NSF award \#2343595. 
Thiruvathukal acknowledges support from NSF award \#2537309. 
\fi

\GKT{Second TODO very important. Double blind does not mean not-fingerprintable.}
\balance 
\bibliographystyle{IEEEtran}
\bibliography{bib/agenthub}

\end{document}